# Attacks on Node Attributes in Graph Neural Networks

**Ying Xu, Michael Lanier, Anindya Sarkar, Yevgeniy Vorobeychik**

Washington University in St. Louis
{x.ying1, lanier.m, anindya, yvorobeychik}@wustl.edu

## Abstract

Graphs are commonly used to model complex networks prevalent in modern social media and literacy applications. Our research investigates the vulnerability of these graphs through the application of feature based adversarial attacks, focusing on both decision-time attacks and poisoning attacks. In contrast to state-of-the-art models like Net Attack and Meta Attack, which target node attributes and graph structure, our study specifically targets node attributes. For our analysis, we utilized the text dataset Hellaswag and graph datasets Cora and CiteSeer, providing a diverse basis for evaluation. Our research reveals that decision-time attacks employing Projected Gradient Descent (PGD) are more effective than poisoning attacks that deviate from the distributions of node embeddings and the implementation of Graph Contrastive Learning strategies. Nonetheless, under certain conditions, these poisoning methods can outperform PGD. This provides insights for graph data security, pinpointing where graph-based models are most vulnerable and thereby informing the development of stronger defense mechanisms against such attacks.

## 1 Introduction

In today's era of hyperconnectivity, digital platforms such as Twitter, Reddit, and YouTube have transcended their roles as mere communication tools, emerging as vital components in our daily social fabric (Newman, Watts, and Strogatz 2002). These platforms, through their advanced algorithms, facilitate the creation of expansive networks of conversations and affiliations, effectively shaping multifaceted social network graphs (Myers et al. 2014). Within these networks, each interaction, each comment, serves as a node—a crucial point of interconnection and influence. Graph models of social networks are often utilized to measure the impact of behavior adverse to user experiences. Graph models may also serve to identify actors in breach of terms and conditions, or identify cells of individuals working in concert to influence the broader network.

In these networks users feature level data is generally derived from users, and so users have large to total control of this data. Broader graph level data, i.e which nodes are connected, which nodes are present, are generally hard to impossible to control without high access to the model builders data. Additionally, these networks often have stochasticlly missing nodes or edges in the training data, making such networks robust to these types of attacks. In these cases adversarial attacks may be sufficiently obvious to inspection, as (Maretto et al. 2020) noted in point cloud attacks. In contrast to existing graph model attacks, which assume the ability to modify the graph, we study here attacks that only modify feature level data. In Figure 1 we have an example of a comment in a social media network being classified to location. Here the user is located in St. Petersburg, Russia. The user can control the language, time of data. A user can modify the location tag via the use of any common geo spoofing tool. The user wishes instead to be classified as being in the St. Louis area.

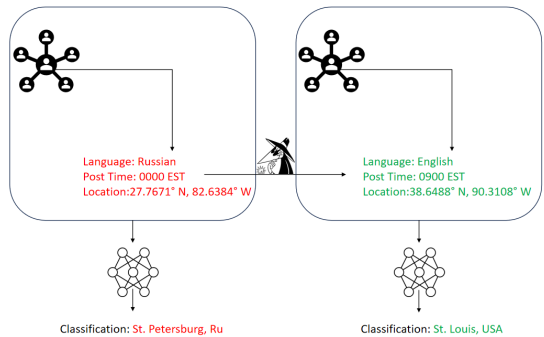

Figure 1: Overview of feature level attack. Graph structure remains fixed.

We examine vulnerability of networks to feature level attacks. We achieve this by employing two categories of adversarial attacks: decision-time attacks and poisoning attacks, as defined in the realm of adversarial machine learning (Vorobeychik and Kantarcioglu 2018). As illustrated in Figure 2, our research pipeline encompasses a comprehensive approach, beginning with text data processing and culminating in the evaluation of adversarial attacks. Distinctively, our methodology concentrates on poisoning attacks that specifically target node attributes, generated from text, a departure from the conventional approaches used in state-of-the-art models such as Meta Attack (Zügner and Günnemann 2019) and Net Attack (Zügner, Akbarnejad, and Günnemann 2018). This targeted strategy aims to streamline the attack process, attempting to match or surpass the efficacy of these more elaborate models. We not only uncover the vulnerabilities inherent in network structures but also recover valuable insights in developing defensive strategies.

**Related Work** The intersection of adversarial learning and Graph Neural Networks (GNNs) is an intriguing space where a model (such as Graph Convolutional Network (GCN) (Kipf and Welling 2016), Graph Attention Network (GAT) (Veličković et al. 2018) and its antagonist engage in a strategic duel. This involves the adversary artfully tweaking the graph structure or node attributes in a bid to manipulate the model's predictions (Zügner, Akbarnejad, and Günnemann 2018).

A key development in this area is NETTACK (Zügner, Akbarnejad, and Günnemann 2018), designed to compromise the structural integrity of graphs. This method conducts poisoning attacks during the training phase of GCN models, subtly disrupting their learning process. In contrast, RL-S2V (Dai et al. 2018) employs reinforcement learning for evasion attacks during testing, altering graph data dynamically to evade detection.

Other approaches, such as those in (Chen et al. 2018) and (Wu et al. 2019), utilize gradient information for refined poisoning strategies. (Chen et al. 2018) iteratively modifies node connections based on significant gradient values, targeting the graph's embedding space. Meanwhile, (Wu et al. 2019) leverages integrated gradients to approximate model gradients, strategically flipping binary values to perturb the graph (Dai et al. 2018; Zügner and Günnemann 2019). Distinct from heuristic-based or direct optimization attacks, Liu et al.'s method (Liu et al. 2023) adopts a curriculum learning approach, progressively generating poisoned graphs starting from basic adversarial knowledge and advancing to more complex strategies.

## 2 Preliminaries

### 2.1 Graph Neural Networks

A graph is typically represented as $G = (V, E)$, where $V$ denotes the set of vertices, and $E$ represents the set of edges. In an undirected graph, the adjacency matrix $A \in \mathbb{R}^{n \times n}$ is symmetric, with $n$ indicating the number of nodes. The degree matrix $D$ is a diagonal matrix where $D_{ii}$ is the degree of node $i$, defined as the sum of the weights of all edges connected to node $i$.

In recent years GNNs have become standard for graph analysis. Among these, two state-of-the-art models GCN and GAT have emerged as the standard for tasks such as node classification, link prediction, and graph classification.

The core of a GCN lies in its propagation rule, defined as:

$$H^{(l+1)} = \sigma \left( \hat{D}^{-\frac{1}{2}} \hat{A} \hat{D}^{-\frac{1}{2}} H^{(l)} W^{(l)} \right) \quad (1)$$

Here, $H^{(l)} \in \mathbb{R}^{n \times d_l}$ denotes the activation matrix at the $l$-th layer, $W^{(l)} \in \mathbb{R}^{d_l \times d_{l+1}}$ is the weight matrix for that layer, $\hat{A} = A + I_n$ is the adjacency matrix augmented with self-loops, and $\hat{D}$ is the degree matrix of $\hat{A}$. The function $\sigma(\cdot)$ represents a nonlinear activation function, typically ReLU.

In contrast, GATs introduce an attention mechanism that governs the aggregation of node features. This mechanism can be formalized as:

$$\alpha_{ij} = \text{softmax}_j \left( \text{LeakyReLU} \left( \vec{a}^T \left[ W\vec{h}_i \parallel W\vec{h}_j \right] \right) \right) \quad (2)$$

$$\vec{h}_i' = \sigma \left( \sum_{j \in \mathcal{N}(i)} \alpha_{ij} W \vec{h}_j \right) \quad (3)$$

Where $\vec{h}_i$ is the feature vector of node $i$, $W$ is a shared linear transformation for all nodes, and $\vec{a}$ is the attention mechanism's weight vector. The LeakyReLU function adds non-linearity, and $\text{softmax}_j$ normalizes the attention coefficients across all choices of $j$. The final feature $\vec{h}_i'$ for node $i$ is derived as a weighted sum of its neighbors' features $\mathcal{N}(i)$, weighted by the attention coefficients $\alpha_{ij}$.

GATs have shown exceptional performance in scenarios where the relative significance of neighboring nodes varies substantially (Trivedi et al. 2019). Their ability to assign different weights to different parts of the graph renders them highly effective and adaptable for a variety of graph-based learning tasks.

### 2.2 Adversarial Machine Learning

Adversarial Machine Learning explores the intersection of machine learning and cybersecurity, focusing on how machine learning systems can be compromised by adversarial actors. Actors craft data to elicit misbehavior from models, either at inference time or in training time. Actors may also extract information about the model or its training data.

For a machine learning model $f : \mathcal{X} \to \mathcal{Y}$, where $\mathcal{X}$ is the input space and $\mathcal{Y}$ is the output space, an adversarial example $\mathbf{x}_{adv}$ is defined as:

$$\mathbf{x}_{adv} = \mathbf{x} + \delta, \text{ such that } f(\mathbf{x}_{adv}) \neq y \quad (4)$$

The perturbation $\delta$ is typically small, often constrained by a $p$-norm (like $L_\infty$ or $L_2$) to ensure the adversarial example remains similar to the original:

$$\|\delta\|_p \leq \epsilon \quad (5)$$

Unlike traditional machine learning methods, which aim to minimize a loss function, adversarial attacks seek to increase the model's loss. Specifically, for an adversarial example, the objective is formulated as maximizing the loss, subject to a constraint on the perturbation magnitude $\delta$. The formal representation is:

$$\max_\delta \mathcal{L}(f(\mathbf{x} + \delta), y) \quad \text{s.t.} \quad \|\delta\|_p \leq \epsilon \quad (6)$$

**Decision Time Attack** Decision-time attacks introduce adversarial examples at inference. Several popular techniques exist including Fast Gradient Sign Method (FGSM) by (Goodfellow, Shlens, and Szegedy 2015) and the Projected Gradient Descent (PGD) method (Madry et al. 2018).

FGSM creates adversarial examples by perturbation $\eta$ to maximize loss under $\infty$-norm constraint $\epsilon$:

$$\eta = \epsilon \cdot \text{sign}(\nabla_x J(\theta, x, y)) \tag{7}$$

PGD, an iterative method, takes steps in the gradient direction, followed by projection to stay within the $\epsilon$-ball:

$$x^{(t+1)} = \Pi_{x+S}\left(x^{(t)} + \alpha \cdot \text{sign}(\nabla_x J(\theta, x^{(t)}, y))\right) \tag{8}$$

where $\Pi$ represents the projection operation and $\alpha$ is the fixed gradient step size.

**Poisoning Attack** Poisoning attacks distinguish themselves from decision-time attacks by targeting the training phase of machine learning models. Unlike decision-time attacks, which manipulate inputs during the inference phase, poisoning attacks alter the training data, thereby compromising the learning process itself. These attacks involve introducing carefully crafted perturbations or entirely fabricated data points into the training set, leading the model to learn incorrect patterns.

Formally, consider a training dataset $\mathcal{D} = \{(\mathbf{x}_i, y_i)\}_{i=1}^N$ where $N$ is the number of training samples. A poisoning attack aims to inject a set of adversarial examples $\mathcal{D}_{adv} = \{(\mathbf{x}_{adv}, y_{adv})\}$ into $\mathcal{D}$, resulting in a poisoned dataset $\mathcal{D}_p = \mathcal{D} \cup \mathcal{D}_{adv}$. The objective is to maximize the model's loss on a separate, untainted test dataset:

$$\max_{\mathcal{D}_{adv}} \mathcal{L}(f_\theta(\mathcal{D}_p), \mathcal{Y}_{test}) \tag{9}$$

where $f_\theta$ represents the learning model parameterized by $\theta$, and $\mathcal{Y}_{test}$ is the set of true labels for the test dataset. The challenge in crafting $\mathcal{D}_{adv}$ lies in ensuring that the adversarial examples are subtly different from genuine data points, making them difficult to detect yet effective in altering the model's learning trajectory.

## 3 Adversarial Attacks

This section outlines the construction of graphs and the methodologies devised to evaluate the effectiveness of our proposed adversarial attacks on GNNs, particularly when applied to text datasets.

### 3.1 Graph Building by Using Text Dataset

Overall, our paper's pipeline is showed in the following figure 2. Our approach to adversarial studies on GNNs begins with preprocessing the text data from the Hellaswag dataset (Zellers et al. 2019). We utilize the BERT model (Devlin et al. 2018) to extract contextual representations from the "$ctx\_a$" text features of the dataset. These representations are then transformed into node embeddings for our graph.

The construction of the graph involves establishing edges based on the relationships between node embeddings. We compute the cosine similarity between each pair of node embeddings to measure the strength of their interconnections. An edge is created between two nodes if their cosine similarity surpasses a predetermined threshold, ensuring connectivity among nodes that share substantial similarity.

For label processing, we employ a LabelEncoder to convert categorical labels into a numerical format, making them compatible with the non-directed graph structure utilized in our analysis. This encoding facilitates the integration of labels into the graph, preparing the dataset for subsequent adversarial attack methodologies.

### 3.2 Adversarial Attacks

**Decision-Time Attack Targeting Top-K Degree Nodes** Our approach extends beyond standard PGD techniques for decision-time attacks by specifically targeting the top-K degree nodes in the graph. This strategy aims to assess the impact on the most influential nodes, hypothesizing that manipulating these nodes could have a more pronounced effect on the overall performance of the model.

The process for executing this targeted decision-time attack is detailed below:

---
Algorithm 1: Targeted Top-K Degree Decision-Time Attack Procedure using PGD

---
**Input:** Graph $G(V, E)$, Node Embeddings $Emb$
**Parameter:** Number of Target Nodes $K$, Number of Iterations $N$, Learning Rate $\alpha$, Perturbation Norm Bound $\epsilon$
**Output:** Modified Graph $G'$
**for** each node $v \in V$ **do**
    Initialize perturbation $\eta_v \leftarrow \mathbf{0}$
**end for**
**for** $i = 1$ to $N$ **do**
    **for** each node $v$ in the top $K$ degree nodes **do**
        Calculate gradient: $\nabla_{\eta_v} J(Emb(v), y)$
        Update perturbation: $\eta_v \leftarrow \eta_v + \alpha \cdot \text{sign}(\nabla_{\eta_v} J)$
        Project $\eta_v$ onto an $\epsilon$-ball: $\eta_v \leftarrow \min(\epsilon, \|\eta_v\|_p) \cdot \frac{\eta_v}{\|\eta_v\|_p}$
        Apply perturbation: $Emb'(v) \leftarrow Emb(v) + \eta_v$
    **end for**
**end for**
**return** Modified Graph $G'$ with updated embeddings $Emb'$

---

This pseudocode outlines the procedure for applying decision-time attacks, utilizing PGD, to the top-K degree nodes. By focusing on these highly connected nodes, the experiment seeks to reveal how such targeted perturbations influence the robustness and performance of GNNs under adversarial conditions.

**Deviating the Distribution of Node Embeddings** In our poisoning attack strategy, a critical step involves altering the original distribution of node embeddings. A naive approach to achieve this is by adding random noise to the node embeddings. To implement a more systematic strategy, we calculate the mean embedding $\bar{\mathbf{e}}$ of the entire node set. We then use a control parameter $\lambda$ to regulate the degree of adjustment applied to each node embedding $\tilde{\mathbf{e}}_i$:

$$\tilde{\mathbf{e}}_i = (1 - \lambda)\mathbf{e}_i + \lambda\bar{\mathbf{e}} \tag{10}$$

Here, $\bar{\mathbf{e}}$ is the centroid of the node embedding space, computed as $\bar{\mathbf{e}} = \frac{1}{n}\sum_{i=1}^n \mathbf{e}_i$. This transformation shifts each node embedding towards the mean, effectively reducing the variance among the embeddings. The intention behind this homogenization is to impair the model's ability to differentiate between various graph structures, potentially diminishing its classification accuracy and overall robustness.

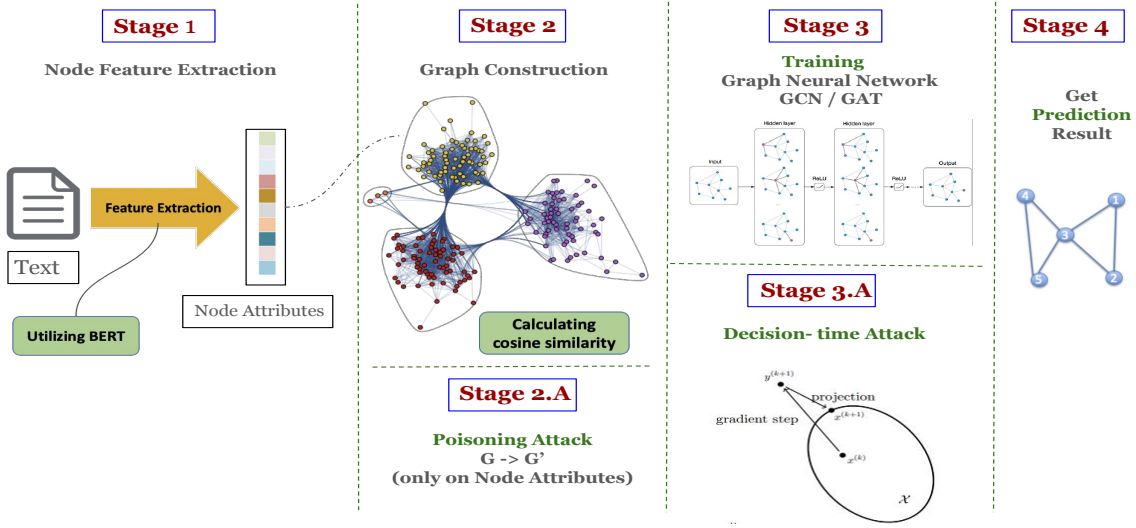

Figure 2: Schematic representation of the research workflow, outlining the progression from text data processing to adversarial attack evaluation.

**Applying Graph Contrastive Learning to Poison Node Embeddings** Graph Contrastive Learning (GCL) can be exploited for poisoning attacks aimed at compromising a GNN's performance (Zhu et al. 2023). We cast this as an optimization problem that targets the node embeddings of the entire graph. The adversarial objective is to find perturbations that, when applied to the node embeddings, hinder the GNN's accuracy. The optimization problem is given by:

$$\min_{E(x^*)} \quad L = \lambda L_{sim} + (1-\lambda) L_{dis} \quad (11)$$

where $L_{sim}$ is the similarity loss between poisoned embeddings, $L_{dis}$ is the dissimilarity loss encouraging differentiation from original embeddings.

The similarity loss $L_{sim}$ is defined as the negative mean similarity between all poisoned embeddings:

$$L_{sim} = -\frac{1}{N} \sum_{i=1}^{N} \sum_{j=1}^{N} \text{sim}(E(x_i^*), E(x_j^*)), \quad (12)$$

and the dissimilarity loss $L_{dis}$ is the log-sum-exp of similarities between original and poisoned embeddings, promoting divergence:

$$L_{dis} = \sum_{i=1}^{N} \log \left( \sum_{j=1}^{N} e^{\text{sim}(E(x_i), E(x_j^*))} \right). \quad (13)$$

In this formulation, $\lambda$ is the hyperparameter that balances the trade-off between similarity and dissimilarity losses, thus guiding the severity of the poisoning.

Applying this framework, we target the node embeddings across the entire graph to induce significant impacts on the GNN's functionality.

---

Algorithm 2: Poisoning Node Embeddings using GCL

**Input:** Graph $G(V, E)$, Node embeddings $Emb$
**Parameters:** Total epochs $T$
**Output:** Poisoned embeddings $Emb^*$
Define the optimizer for the poisoning process
**for** $epoch = 1$ to $T$ **do**
  Calculate the similarity matrix for all node embeddings
  Define similarity loss $L_{sim}$ and dissimilarity loss $L_{dis}$ to maximize dissimilarity from original embeddings.
  Compute the total loss as a weighted sum of $L_{sim}$ and $L_{dis}$
  Apply perturbations to embeddings $Emb$ to get poisoned embeddings $Emb^*$ as follows:

$$E(x^*) \leftarrow \min_{E(x^*)} [\lambda L_{sim} + (1-\lambda) L_{dis}]$$

**end for**
**return** $Emb^* = E(x^*)$

---

This pseudocode outlines the steps to apply GCL for poisoning the embeddings of the entire graph. The process involves iteratively updating the node embeddings to solve the defined optimization problem, ensuring the perturbed embeddings $Emb^*$ enforce the GNN's node prediction to be wrong.

## 4 Experiments

This section delineates the experimental results from deploying Graph Neural Network (GNN) models on datasets under normal conditions and those subjected to adversarial attacks, including both decision-time and poisoning strategies. We specifically examine the performance degradation when exposed to these white-box attacks.

### 4.1 Model Settings

In our experiments, we utilize two distinct GNN architectures: GCN(Kipf and Welling 2016) and GAT(Veličković et al. 2018). Both are designed to handle graph-structured data, but they differ in their approach to feature extraction and node aggregation. For both network structures, as detailed in Tables 1 and 2, the input is configured as $[768, 1, 1] \times N$ for text datasets, representing the initial feature set after feature extractions. The models are tasked with outputting class predictions corresponding to the number of classes (num_classes) in the classification task.

Table 1: GCN Architecture

| Layers | Configuration |
|---|---|
| Input | Text (Graph Data) |
| Feat. Extraction | BERT (Initial Feature Representation) |
| Conv1 | GCNConv(num_features, 256) |
| ReLU1 | ReLU Activation |
| Dropout1 | Dropout (Training Phase) |
| Conv2 | GCNConv(256, 32) |
| ReLU2 | ReLU Activation |
| Dropout2 | Dropout (Training Phase) |
| Conv3 | GCNConv(32, num_classes) |
| Output | Logits (num_classes) |

Table 2: GAT Architecture

| Layers | Configuration |
|---|---|
| Input | Graph Data (Nodes Features and Edge Index) |
| Conv1 | GATConv(num_features, 256) |
| Activation1 | LeakyReLU Activation |
| Dropout1 | Dropout (Training Phase) |
| Conv2 | GATConv(256, 32) |
| Activation2 | LeakyReLU Activation |
| Dropout2 | Dropout (Training Phase) |
| Conv3 | GATConv(32, num_classes) |
| Output | Logits (num_classes) |

### 4.2 Clean Dataset Result

We begin by evaluating the performance of our models on clean datasets, free from any adversarial modifications. The datasets used for this assessment include the text-based Hellaswag dataset and the graph datasets Cora and CiteSeer. This evaluation serves as a baseline to understand the models' behavior under normal conditions.

### 4.3 Decision-Time Attack

In our decision-time attack experiments, we utilized PGD with various norms - $L_1$, $L_2$, and $L_\infty$ within an epsilon-ball to bound the changes in the dataset. The impact of

Table 3: Performance comparison of clean models on multiple datasets.

| Dataset | Model | Validation Acc | Testing Acc |
|---|---|---|---|
| Hellaswag | GCN | 93.75% | 92.42% |
| Hellaswag | GAT | 92.19% | 93.94% |
| Cora | GCN | 92.19% | 93.94% |
| Cora | GAT | 77.60% | 80.10% |
| CiteSeer | GCN | 64.60% | 65.30% |
| CiteSeer | GAT | 64.80% | 65.80% |

these attacks on the final accuracy rate can be influenced by two main factors: the choice of norm and the value of $\epsilon$. A smaller norm restricts the perturbation boundary, resulting in less perturbation, while an increase in $\epsilon$ allows for greater perturbation. Below, we present the results showing how the accuracy rate varies with different $\epsilon = 0.1$ values for both the Hellaswag and Cora datasets:

Table 4: Test Loss and Accuracy for Different Norms with $\epsilon = 0.1$ on Hellaswag and Cora Datasets

| Dataset | Norm | Test Loss | Test Accuracy |
|---|---|---|---|
| Hellaswag | $L_1$ | 0.1223 | 95.4545% |
| Hellaswag | $L_2$ | 0.3755 | 83.3333% |
| Hellaswag | $L_\infty$ | 1.3701 | 30.3030% |
| Cora | $L_1$ | 0.7637 | 77.3% |
| Cora | $L_2$ | 12.7102 | 1.0% |
| Cora | $L_\infty$ | 304.5575 | 0.9% |

After examining the impact of different norms, we further explored the influence of varying epsilon $\epsilon$ values. As $\epsilon$ increases, test accuracy initially drops significantly, as is typically observed. However, beyond a certain point, the accuracy tends to stabilize, suggesting an inherent level of robustness in the models against adversarial perturbations.

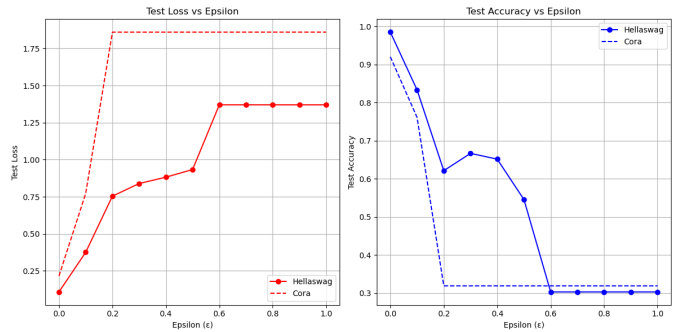

Figure 3: Comparison of Test Loss vs Epsilon for Hellaswag and Cora datasets using the $L_2$ norm.

The numerical results for different epsilon $\epsilon$ values are summarized in the following table 5:

Table 5: Test Loss and Accuracy for Different Epsilon $\epsilon$ Values using $L_2$ Norm

| Epsilon $\epsilon$ | Hellaswag | | Cora | |
|---|---|---|---|---|
| | Test Loss | Accuracy | Test Loss | Accuracy |
| 0.0 | 0.11 | 98.48% | 0.22 | 92.00% |
| 0.1 | 0.38 | 83.33% | 0.77 | 76.10% |
| 0.2 | 0.75 | 62.12% | 1.86 | 31.90% |
| 0.3 | 0.84 | 66.67% | 1.86 | 31.90% |
| 0.4 | 0.88 | 65.15% | 1.86 | 31.90% |
| 0.5 | 0.93 | 54.55% | 1.86 | 31.90% |
| 0.6 | 1.37 | 30.30% | 1.86 | 31.90% |
| 0.7 | 1.37 | 30.30% | 1.86 | 31.90% |
| 0.8 | 1.37 | 30.30% | 1.86 | 31.90% |
| 0.9 | 1.37 | 30.30% | 1.86 | 31.90% |
| 1.0 | 1.37 | 30.30% | 1.86 | 31.90% |

### 4.4 Poisoning Attack

**Net Attack** In our study, we examine the impact of the Net Attack (Zügner, Akbarnejad, and Günnemann 2018), a strategy that manipulates both the features and structure of a graph. The attack is constrained by $|A^{(t)} - A^{(0)}| + |X^{(t)} - X^{(0)}| < \Delta$, where $A$ represents the adjacency matrix and $X$ represents the node attributes. This constraint ensures that the perturbations remain subtle, yet they are potent enough to disrupt the graph model's performance effectively.

Our results indicate varying levels of impact based on the type of perturbation applied. Isolated feature perturbations show a relatively modest effect on model performance, as seen in Figure 4. This observation suggests that feature manipulation alone may not sufficiently compromise a robust graph model.

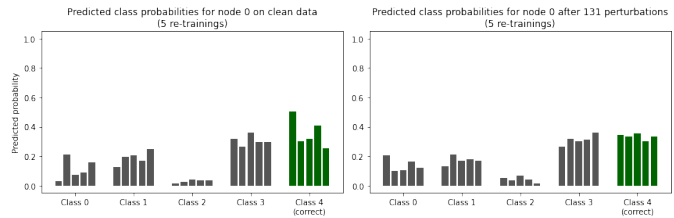

Figure 4: Results for Net Attack when only perturbing features.

In contrast, the application of both feature and structural perturbations results in a more pronounced degradation of model performance, as demonstrated in Figure 5. This highlights the effectiveness of combined feature and structural alterations in poisoning attacks.

These findings underscore the need for considering both feature and structural elements in designing effective poisoning strategies for graph neural networks. Additionally, they shed light on potential vulnerabilities in these models, offering insights for enhancing their resilience against adversarial manipulations.

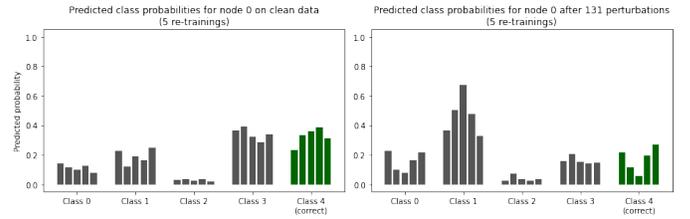

Figure 5: Results for Net Attack when perturbing both features and structure.

**Meta Attack** The Meta Attack algorithm (Zügner and Günnemann 2019) conceptualizes the graph structure matrix as a hyperparameter. The key process involves computing the gradient of the attacker's loss with respect to this matrix after training, denoted as $\nabla_{G^{\text{meta}}}$. This is mathematically represented as:

$$\nabla_{G^{\text{meta}}} := \nabla_G L_{\text{atk}}(f_{\theta^*}(G)),$$

subject to the condition that $\theta^* = \text{opt}_\theta(L_{\text{train}}(f_\theta(G)))$, where $L_{\text{train}}$ and $L_{\text{atk}}$ are the training and attack loss functions, respectively.

This method combines training loss and self-training loss, modulated by the parameter $\lambda$, to effectively simulate a poisoning attack on GNNs. It highlights the criticality of the graph structure in the resilience of GNNs against sophisticated adversarial techniques.

Intriguingly, when applying the Meta Attack to the same text dataset used for the Net Attack, the outcomes were surprisingly akin. Contrary to initial assumptions, the Meta Attack did not markedly alter the dataset, yielding results nearly identical to the original. This unexpected result prompts further investigation into the underlying reasons why this attack strategy might inadvertently enhance model performance, rather than degrade it.

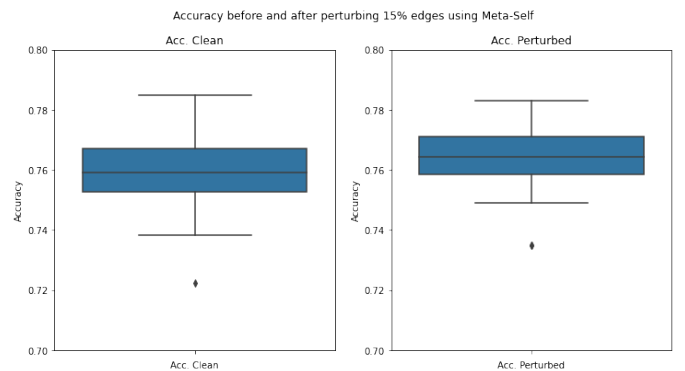

Figure 6: Meta Attack Result applied on the text dataset "Hellaswag".

*The outcomes of the Meta Attack and Net Attack, as indicated above, demonstrate limited effectiveness when the graph structure is kept constant and only the node attributes are targeted.*

**Deviating Distributions of Node Embeddings** In our examination of Mean Node Embedding Poisoning Attacks, we manipulated the lambda $\lambda$ parameter to control the extent of deviation in node embeddings. This experiment, conducted on the Cora dataset, aimed to understand how varying levels of embedding deviation impact the model's performance. The results, which include measures of test loss and accuracy across a range of $\lambda$ values, are visually represented in Figure 7 and quantitatively summarized in the table below.

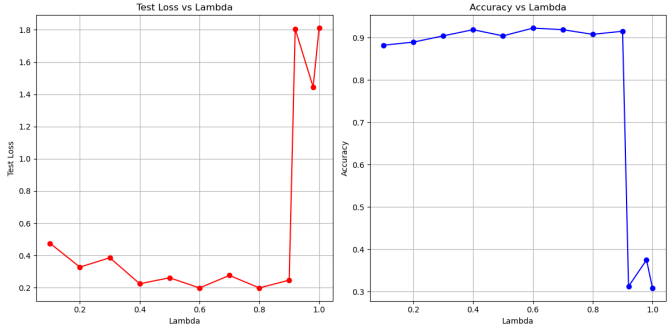

Figure 7: Impact of varying lambda on embeddings in the Cora dataset.

The table presents a detailed breakdown of the test loss and accuracy corresponding to each $\lambda$ value, providing insights into how the degree of embedding deviation influences the model's effectiveness in classifying graph data.

Table 6: Test Loss and Accuracy for Different Lambda Values on Cora Dataset

| Lambda | Test Loss | Accuracy |
| --- | --- | --- |
| 0.1 | 0.4756 | 88.2353% |
| 0.2 | 0.3278 | 88.9706% |
| 0.3 | 0.3860 | 90.4412% |
| 0.4 | 0.2247 | 91.9118% |
| 0.5 | 0.2615 | 90.4412% |
| 0.6 | 0.1986 | 92.2794% |
| 0.7 | 0.2764 | 91.9118% |
| 0.8 | 0.1988 | 90.8088% |
| 0.9 | 0.2469 | 91.5441% |
| 0.92 | 1.8055 | 31.2500% |
| 0.98 | 1.4456 | 37.5000% |
| 1.0 | 1.8109 | 30.8824% |

These findings highlight the sensitivity of graph-based models to alterations in node embeddings and underscore the importance of carefully calibrating embedding-related parameters in the face of potential adversarial manipulations.

**GCL Poisoning Attack** In our experiment, we explored various combinations of the $\lambda$ parameter and iteration counts across three datasets. The comprehensive results are presented in Table 7, including comparisons between clean data and two poisoning attack methods. As depicted in Figure 8, we observed a significant decline in the model's performance when subjected to poisoned datasets, with accuracy plummeting to 33.00%. In contrast, the accuracy of the model on the validation dataset eventually stabilized at around 90%. This suggests that while the poisoning attack—comparable to a backdoor attack as described in (Zhang et al. 2020; Gu et al. 2023)—significantly diminishes performance on poisoned data, it has a relatively minor impact on clean datasets.

| Dataset | GNNs | Clean Data | Deviate Dist | GCL |
| --- | --- | --- | --- | --- |
| Hellaswag | GCN | 92.42% | 28.24% | 56.06% |
| Hellaswag | GAT | 93.94% | 14.50% | 54.55% |
| Cora | GCN | 93.94% | 31.25% | 31.90% |
| Cora | GAT | 80.10% | 31.31% | 33.00% |
| CiteSeer | GCN | 65.30% | 22.33% | 30.80% |
| CiteSeer | GAT | 65.80% | 17.84% | 19.60% |

Table 7: The result of different kinds of poisoning attacks

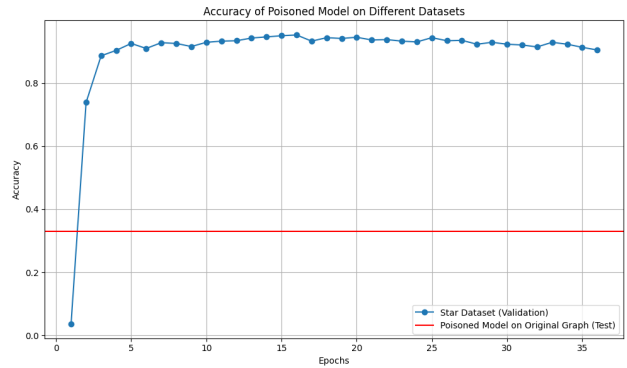

Figure 8: Performance comparison of the model on the original dataset before and after the poisoning attack.

## 5 Conclusion and Future Work

This study addressed an issue in the realm of graph datasets: executing attacks using node attributes derived from textual data, a scenario that holds particular significance in social network analysis, and citation networks. Our experimental framework included both the text dataset: Hellaswag and graph datasets: Cora and CiteSeer. We comprehensively explored both decision-time and poisoning attacks, with a specific focus on the latter. Our approach diverged from standard methods, such as those in Meta Attack(Zügner and Günnemann 2019) and Net Attack(Zügner, Akbarnejad, and Günnemann 2018), by exclusively altering node embeddings without affecting the graph structure and node attributes simultaneously. The results from our investigation demonstrate that, even within these constraints, it is possible to effectively undermine graph analysis based on neural network architectures.

Notably, we observed that decision-time attacks exhibit a more pronounced impact compared to poisoning attacks.

This could be attributed to the direct application of decision-time attacks on datasets, in contrast to poisoning attacks, which target models indirectly. Consequently, decision-time attacks can significantly alter results, highlighting their potency.

However, our findings also suggest that to achieve more substantial adversarial effects, stronger attack constraints might be necessary. This includes implementing higher values of $\epsilon$ or $\lambda$. We speculate that the resilience seen in our experiments may be partially due to the informational content inherent in the links of the graph, potentially offsetting the node embedding perturbations to some degree.

This research contributes to a deeper understanding of adversarial vulnerabilities in graph-based models, especially those using text-derived node embeddings, and lays the groundwork for further exploration in this important area of adversarial machine learning. Next steps may include:

1. *Textual Embedding Revision:* Another intriguing avenue involves revising embeddings to enable backpropagation to the original text. This approach would assess if attacks adversely impact the text dataset. The challenge lies in converting tensor data back into coherent text, akin to image restoration techniques used in models like Clip (Radford et al. 2021), but specifically tailored for textual content.

2. *Exploring Defensive Strategies:* Additionally, investigating defensive mechanisms against these attacks could shed light on the robustness of graph-based models. This includes developing techniques to detect and mitigate adversarial modifications without compromising the model's performance on genuine data.

This research lays the groundwork for further exploration into the vulnerabilities of graph-based models, especially those utilizing text-derived embeddings, and opens new frontiers for both offensive and defensive strategies in adversarial machine learning.

# References

Chen, J.; Wu, Y.; Xu, X.; Chen, Y.; Zheng, H.; and Xuan, Q. 2018. Fast gradient attack on network embedding. *arXiv preprint arXiv:1809.02797*.

Dai, H.; Li, H.; Tian, T.; Huang, X.; Wang, L.; Zhu, J.; and Song, L. 2018. Adversarial attack on graph structured data. In *International conference on machine learning*, 1115–1124. PMLR.

Devlin, J.; Chang, M.; Lee, K.; and Toutanova, K. 2018. BERT: Pre-training of Deep Bidirectional Transformers for Language Understanding. *CoRR*, abs/1810.04805.

Goodfellow, I.; Shlens, J.; and Szegedy, C. 2015. Explaining and Harnessing Adversarial Examples. In *International Conference on Learning Representations*.

Gu, C.; Zheng, X.; Xu, J.; Wu, M.; Zhang, C.; Huang, C.; Cai, H.; and Huang, X. 2023. Watermarking PLMs on Classification Tasks by Combining Contrastive Learning with Weight Perturbation. In *Conference on Empirical Methods in Natural Language Processing*.

Kipf, T. N.; and Welling, M. 2016. Semi-supervised classification with graph convolutional networks. *arXiv preprint arXiv:1609.02907*.

Liu, H.; Zhao, P.; Xu, T.; Bian, Y.; Huang, J.; Zhu, Y.; and Mu, Y. 2023. Curriculum Graph Poisoning. In *Proceedings of the ACM Web Conference 2023*, WWW '23, 2011–2021. New York, NY, USA: Association for Computing Machinery. ISBN 9781450394161.

Madry, A.; Makelov, A.; Schmidt, L.; Tsipras, D.; and Vladu, A. 2018. Towards Deep Learning Models Resistant to Adversarial Attacks. In *6th International Conference on Learning Representations, ICLR 2018, Vancouver, BC, Canada, April 30 - May 3, 2018, Conference Track Proceedings*.

Maretto, R. V.; Fonseca, L. M. G.; Jacobs, N. B.; Körting, T. S.; Bendini, H. N.; and Parente, L. L. 2020. Spatio-Temporal Deep Learning Approach to Map Deforestation in Amazon Rainforest. *IEEE Geoscience and Remote Sensing Letters*, 18(5): 771–775. Impact factor: 3.534.

Myers, S. A.; Sharma, A.; Gupta, P.; and Lin, J. 2014. Information network or social network? The structure of the Twitter follow graph. In *Proceedings of the 23rd International Conference on World Wide Web*, 493–498.

Newman, M. E.; Watts, D. J.; and Strogatz, S. H. 2002. Random graph models of social networks. *Proceedings of the national academy of sciences*, 99(suppl_1): 2566–2572.

Radford, A.; Kim, J. W.; Hallacy, C.; Ramesh, A.; Goh, G.; Agarwal, S.; Sastry, G.; Askell, A.; Mishkin, P.; Clark, J.; et al. 2021. Learning transferable visual models from natural language supervision. In *International conference on machine learning*, 8748–8763. PMLR.

Trivedi, R.; Farajtabar, M.; Biswal, P.; and Zha, H. 2019. Dyrep: Learning representations over dynamic graphs. In *International conference on learning representations*.

Veličković, P.; Cucurull, G.; Casanova, A.; Romero, A.; Liò, P.; and Bengio, Y. 2018. Graph Attention Networks. arXiv:1710.10903.

Vorobeychik, Y.; and Kantarcioglu, M. 2018. *Adversarial machine learning / Yevgeniy Vorobeychik, Murat Kantarcioglu*. Synthesis lectures on artificial intelligence and machine learning, 38. San Rafael, California: Morgan Claypool. ISBN 9781681733968.

Wu, H.; Wang, C.; Tyshetskiy, Y.; Docherty, A.; Lu, K.; and Zhu, L. 2019. Adversarial examples on graph data: Deep insights into attack and defense. *arXiv preprint arXiv:1903.01610*.

Zellers, R.; Holtzman, A.; Bisk, Y.; Farhadi, A.; and Choi, Y. 2019. HellaSwag: Can a machine really finish your sentence? *arXiv preprint arXiv:1905.07830*.

Zhang, Z.; Jia, J.; Wang, B.; and Gong, N. Z. 2020. Backdoor Attacks to Graph Neural Networks. *CoRR*, abs/2006.11165.

Zhu, Y.; Ai, X.; Vorobeychik, Y.; and Zhou, K. 2023. Homophily-Driven Sanitation View for Robust Graph Contrastive Learning. arXiv:2307.12555.

Zügner, D.; Akbarnejad, A.; and Günnemann, S. 2018. Adversarial Attacks on Neural Networks for Graph Data. *Proceedings of the 24th ACM SIGKDD International Conference on Knowledge Discovery & Data Mining*, 2847–2856.

Zügner, D.; and Günnemann, S. 2019. Adversarial Attacks on Graph Neural Networks via Meta Learning. In *International Conference on Learning Representations*.